# The Impact of Dispersion on Amplitude and Frequency Noise in a Yb-fiber Laser Comb

Lora Nugent-Glandorf,[1]* Todd A. Johnson,[1] Yohei Kobayashi,[2] Scott A. Diddams[1]

[1]*NIST Time and Frequency Division, Mail Stop 847, 325 Broadway, Boulder, CO 80305 USA*
[2]*The University of Tokyo Institute for Solid State Physics, Kashiwanoha 5-1-5, Kashiwa, Chiba 277-8581 Japan*
*Corresponding authors: LNG@boulder.nist.gov, scott.diddams@nist.gov*

We describe a Yb-fiber based laser comb, with a focus on the relationship between net-cavity dispersion and the frequency noise on the comb. While tuning the net cavity dispersion from anomalous to normal, we measure the amplitude noise (RIN), offset frequency ($f_{CEO}$) linewidth, and the resulting frequency noise spectrum on $f_{CEO}$. We find that the laser operating at zero net-cavity dispersion has many advantages, including an approximately 100x reduction in free-running $f_{CEO}$ linewidth and frequency noise power spectral density between laser operation at normal and zero dispersion. In this latter regime, we demonstrate a phase-locked $f_{CEO}$ beat with low residual noise.

Yb-doped, fiber femtosecond laser combs have been the subject of recent attention due to their high efficiency, high repetition rate capabilities, low noise, and ease of direct diode pumping and amplification (1-4). Yb-fiber combs have proven particularly useful in spectroscopic applications, where amplification and nonlinear frequency generation can push the initial 1.03 μm wavelength into the mid-IR (5,6) or ultraviolet (7). There are several fiber laser designs (8) that can operate in various group delay dispersion (GDD) regimes, from soliton-like (net GDD is negative) to similariton pulse propagation (net GDD is normal). In all of these designs, net-cavity dispersion plays a significant role in the laser dynamics and must be carefully managed.

We employ a laser design that can be tuned readily across a range of net-cavity dispersion, in order to explore the relationship between dispersion and noise on the frequency comb (both amplitude and frequency). Since a frequency comb with low amplitude and frequency noise is necessary for precision spectroscopy and other applications, it is important to understand the properties of these Yb-fiber laser combs in all dispersion regimes. Noise on the comb teeth can originate from many different sources inside the cavity, including amplitude noise from the pump diode, amplified stimulated emission (ASE) jitter, cavity loss, and cavity length changes (9). With a pair of volume holographic gratings, we are able to adjust the dispersion inside the cavity in order to study the dependence of frequency comb noise on net cavity dispersion, and to identify the optimum dispersion for obtaining a low-noise comb with narrow linewidth carrier-envelope offset frequency ($f_{CEO}$). In the time domain picture of frequency combs, $f_{CEO}$ is the product of the repetition rate and the carrier-envelope phase ($\phi_{ce}$) change per cavity round trip. Noise on the repetition rate, or 'timing jitter', therefore, is directly coupled to noise on $f_{CEO}$ (10-12). The width of $f_{CEO}$, while not the complete picture of noise in the laser system, is a sensitive indicator of frequency noise when tuning the dispersion in the laser.

We have designed and built the Yb-fiber laser shown in Fig. 1 (1). A 976 nm fiber pigtailed laser diode provides pump light at 350 to 450 mW average power and is spliced to a wavelength division multiplexer (WDM). The output side of the WDM contains 70 cm of Yb gain fiber between two sections of single mode fiber (SMF), while the input side is only SMF. The free-space section contains a polarizing beam splitter for output coupling, an isolator that promotes lasing in the correct direction, and a double-pass through two volume holographic transmission gratings which provides anomalous dispersion compensation for the ~1.5 m total fiber path (normal dispersion). The second grating can be translated a total of 1 cm, corresponding to a dispersion change of 16,000 fs$^2$ at a center wavelength of 1030 nm. Mode-locking is initiated by careful movements of the three free-space waveplates. Depending on mode-locking conditions, the laser output power ranges from 50 to 100 mW at a 100 MHz repetition rate. As is seen in the inset of Fig. 1, the optical spectrum varies considerably when changing the grating separation to tune the net cavity dispersion of the laser from normal to anomalous.

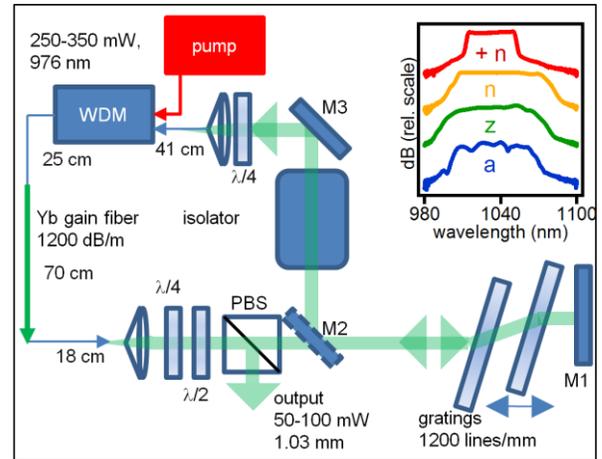

**Fig. 1**: Schematic of Yb-fiber oscillator. WDM=Wavelength Division Multiplexer, PBS=Polarizing Beam Splitter, M1-3=silver mirrors (M2 is a half mirror). *Inset:* Optical spectra at four distinct dispersion regimes in the cavity (+n=more normal, n=normal, z=zero, a=anomalous).

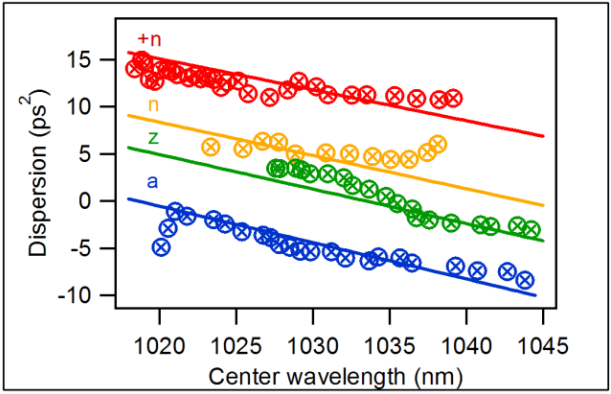

**Fig. 2:** Experimental measurement of the laser cavity GDD (circles); calculation of dispersion as a sum of all the cavity elements (solid lines).

To understand the effects of cavity dispersion on the laser operation, it is first necessary to understand specifically what dispersion regimes are accessible. Knowing the actual cavity dispersion is not straightforward, because fiber dispersion curves (both Yb-doped and SMF) can vary significantly from manufacturing specifications, and from fiber to fiber. Therefore, we first measured the cavity dispersion *in situ* using the technique of Knox, et. al (13). This method is particularly useful, as it measures complete cavity dispersion during mode-locked laser operation, including any nonlinear dispersion arising from self-phase modulation. By putting a slit that is narrower than the beam profile in front of the retro-reflection mirror (M1 in Fig. 1), the optical spectrum, and thus the central wavelength of the mode-locked laser, can be scanned. The width of the mode-locked spectrum remains no less than 75% of the un-filtered spectrum. As the slit is scanned, the free-running repetition rate ($f_{rep}$) is monitored with a photodiode and recorded with a frequency counter. Differentiating the group delay, $T_g$ (= $1/f_{rep}$) vs. the center frequency of the optical spectra ($\nu_0$) yields the net-cavity group-delay dispersion, or GDD. The experimental results for four grating separations are shown in Fig. 2 (circles). Solid lines in Fig. 2 are calculated dispersion based on the sum of the elements inside the cavity, including reasonable estimates of fiber dispersion (Yb fiber estimate from (14) and SMF from Flex1060 fiber data), dispersion from high refractive index materials (glass optics and Terbium Gallium Garnet (15), a material found in optical isolators), and grating dispersion (16). The results show that the four mode-locked spectra in the inset of Fig. 1 correspond to two normal dispersion, one near-zero dispersion, and one anomalous dispersion regime. While the 3rd order dispersion is evident in the slope of the lines in Fig. 2, we define a specific mode-locked spectrum in the range of normal to anomalous by the value of the 2nd order dispersion at the center wavelength, nominally 1035 nm for all the spectra in Fig. 1.

With a clear understanding of the cavity dispersion, we measured the free-running offset frequency, $f_{CEO}$, of the laser operating in several dispersion modes. To this end, we first amplified the chirped oscillator output in a Yb-doped fiber amplifier then recompressed the pulses with an external double pass grating pair (average power ~400 mW, 80 fs pulse duration). Continuum generation was achieved with a 15 cm piece of non-linear suspended-core fiber (17). An $f_{CEO}$ beat was then measured with a standard f-2f interferometer (18). Two sample $f_{CEO}$ beat signals are shown in Fig. 3(a) for the normal and zero dispersion regimes, illustrating the significant difference in linewidth. Similar effects have been observed in other Er and Yb laser systems (19); however, to our knowledge this dependence has not been mapped over a range of cavity dispersions.

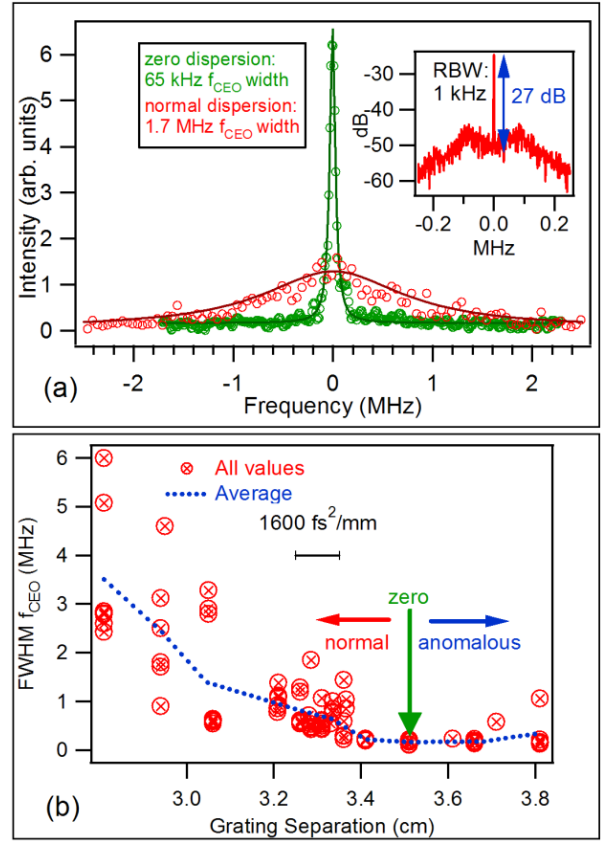

**Fig. 3:** (a) Free-running $f_{CEO}$ beat at normal and zero dispersion points with Lorentzian fit FWHM as the linewidth; *Inset:* Locked $f_{CEO}$ beat (centered at 0 for clarity). (b) Collection of $f_{CEO}$ linewidths vs. net-cavity dispersion (measured by grating separation). Minimum $f_{CEO}$ linewidth at zero dispersion is on the order of 50 kHz.

A complete set of measurements of the free-running $f_{CEO}$ linewidth versus dispersion is given in Figure 3(b). A clear trend of decreasing $f_{CEO}$ linewidth towards zero dispersion (from both the normal and anomalous sides) is seen. For each dispersion regime (i.e., a fixed grating separation) the $f_{CEO}$ linewidth can also change dramatically with pump power (which directly affects the internal cavity circulating power and the spectral width of the pulse). At each grating separation, the pump power was tuned within the full range of stable mode-locking (i.e., no cw breakthrough). Many factors, including the pumping rate, cavity loss and the laser's nonlinear response can impact the $f_{CEO}$ linewidth, including the "turning points" (20) of the $f_{CEO}$ beat and polarization settings. Nevertheless, in the normal dispersion mode we never obtained an $f_{CEO}$ linewidth as narrow as in the zero dispersion regime.

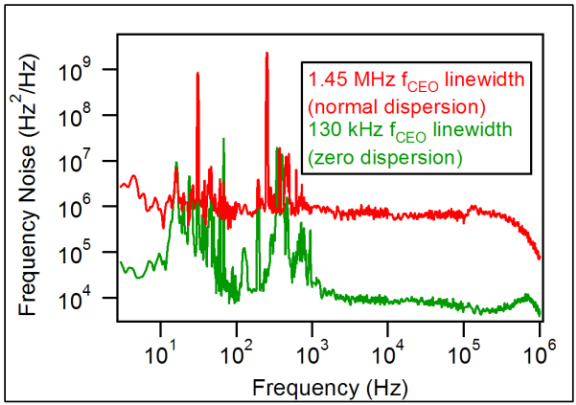

**Fig. 4:** Frequency noise on two sample $f_{CEO}$ beat signals (division factor included to give true $f_{CEO}$ noise).

As the microwave power spectrum (Fig. 3) can ambiguously mix amplitude and frequency noise, we also directly measured the frequency noise spectrum of $f_{CEO}$. This was accomplished by sending the $f_{CEO}$ beat signal into a frequency divider followed by a delay line discriminator and a vector signal analyzer. These results are given in Figure 4 for laser operation at normal and zero dispersion, corresponding to measured free-running $f_{CEO}$ widths of 1.45 MHz and 130 kHz, respectively. The broad $f_{CEO}$ beat clearly has considerably higher FM noise. Calculation of rf spectra from the frequency noise spectra in Fig. 4 (using numerical methods) yields a FWHM of 1.50 MHz (normal) and 104 kHz (zero), a correspondingly 3% and 20% deviation from the experimental values. This might suggest a small contribution from AM noise in the zero dispersion regime, while the much higher FM noise in the normal regime masks any AM contribution. Nonetheless, it is clear that the frequency noise dominates the $f_{CEO}$ linewidths. It is also possible to tightly lock the $f_{CEO}$ beat in the zero dispersion regime as shown in the inset of Fig. 3(a), where the coherent signature on a un-divided copy of $f_{CEO}$ (at 226 MHz) is seen. In Fig. 3(a), 90 % of the power is located within the coherent carrier.

In most frequency combs, there is a strong coupling between amplitude and frequency noise. The amplitude noise of the laser oscillator was measured (Fig. 5) at the same four net-cavity dispersion regimes shown in Figs. 1 and 2. Amplitude noise was measured by monitoring a portion of the laser output with a fast photodiode (1 GHz bandwidth) and using a vector signal analyzer to obtain a power spectral density of the relative intensity noise (RIN=$[\Delta P_{rms}/P_0]^2$ per Hz bandwidth, where P is the optical power). We observed significant differences in the RIN as a function of net cavity dispersion, with zero net cavity dispersion having the lowest RIN. It is not possible to state that 'all other parameters are equal' except the dispersion when comparing the different laser operating modes. Each dispersion point requires adjustment of the pump diode power and waveplate settings in order to achieve stable mode-locking, leading to differences in output coupling and cavity Q. The dispersion is also likely coupled to the loss in the cavity through index of refraction effects for different optical spectra. However, practically, the zero dispersion mode consistently shows the lowest RIN.

Experimentally, the Yb-fiber oscillator presented here operates in its quietest mode near zero dispersion, including reduced amplitude noise on the oscillator, narrowest $f_{CEO}$ linewidth, and quieter frequency and amplitude noise on $f_{CEO}$ itself. This is consistent with the prediction of minimum timing jitter for stretched-pulse fiber lasers near zero net cavity dispersion (9) (21). As mentioned earlier, timing jitter (or noise on $f_{rep}$) is directly coupled to $f_{CEO}$ through the expression $f_{CEO}=(\Delta\phi_{ce}\times f_{rep})/2\pi$. Beyond technical perturbations, the contribution of fundamental quantum noise (ASE) to the circulating power inside the laser directly results in timing jitter. Additionally, there is an indirect effect where ASE-induced frequency jitter is coupled to timing jitter through the group velocity dispersion (9) (21). Thus, minimizing net cavity dispersion also minimizes such jitter, resulting in a narrower $f_{CEO}$ linewidth.

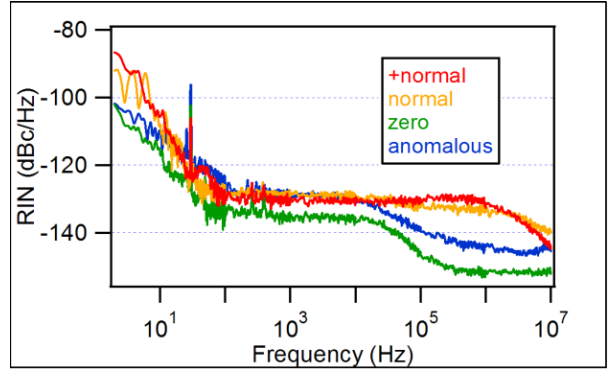

**Fig. 5:** Amplitude noise on the laser output at four dispersion regimes.

In summary, we have experimentally verified the importance of low dispersion for the quietest mode of operation of a Yb fiber laser comb. We give a basis for understanding the aspects of the complex interplay of $f_{CEO}$ linewidth, AM and FM noise, and comb stability.

The authors thank IMRA America, Inc. for the use of the suspended core non-linear fiber (17). This research was funded in part by the U.S. Department of Homeland Security's Science and Technology Directorate through the National Institute of Standards and Technology. We further thank Ingmar Hartl, Nate Newbury, Frank Quinlan and Alex Zolot for helpful comments. Any mention of commercial products does not constitute an endorsement by NIST.